# Femtosecond-laser-pulse characterization and optimization for CARS microscopy

**Vincenzo Piazza[1,*] , Giuseppe de Vito,[1,2] Elmira Farrokhtakin,[3] Gianni Ciofani,[3,4] Virgilio Mattoli[3]**

[1]*Center for Nanotechnology Innovation @NEST, Istituto Italiano di Tecnologia, Piazza San Silvestro 12, I-56127 Pisa, Italy*

[2]*NEST, Scuola Normale Superiore, Piazza San Silvestro 12, I-56127 Pisa, Italy*

[3] *Istituto Italiano di Tecnologia, Center for Micro-BioRobotics @SSSA, Viale RinaldoPiaggio 34, I-56025 Pontedera, Pisa, Italy*

[4]*Polytechnic University of Torino, Department of Aerospace and Mechanical Engineering, Corso Duca degli Abruzzi 24, I-10129 Torino, Italy*

[*] Corresponding author. vincenzo.piazza@iit.it (VP)

## Abstract

We present a simple method and its experimental implementation to completely determine the characteristics of the pump-and-probe pulse and the Stokes pulse in a coherent anti-Stokes Raman scattering microscope at sample level without additional autocorrelators. Our approach exploits the delay line, ubiquitous in such microscopes, to perform a convolution of the pump-and-probe and Stokes pulses as a function of their relative delay and it is based on the detection of the photons emitted from an appropriate non-linear sample. The analysis of the non-resonant four-wave-mixing and sum-frequency-generation signals allows retrieving the pulse duration on the sample and the chirp of each pulse. This knowledge is crucial in maximizing the spectral-resolution and contrast in CARS imaging.

## Introduction

In the coherent-anti-Stokes-Raman-scattering (CARS) process a pair of photons ("pump" and "Stokes") is exploited to excite selected vibrational levels of a sample by choosing them so that their frequency difference matches the vibrational frequencies of interest. A third photon, from the same source as the pump in the so-called two-color CARS implementations, coherently probes the phonon population of the vibrational mode generating strong anti-Stokes emission [1]. This approach is being successfully used to obtain video-rate label-free imaging of biological samples with chemical selectivity [2] and to probe the orientation of chemical bonds in biological samples [3-6]. The third-order nature of the process suggests using ultrashort laser pulses to maximize the generated signal intensity at constant average power delivered to the sample. Conversely, reducing the pulse durations below the lifetime of the Raman modes under investigation (~1 ps in typical biological samples) yields reduced contrast due to non-resonant four-wave-mixing generation as well as poor spectral selectivity. To overcome this issue, ps laser sources are routinely used. On the other hand, these sources are not particularly efficient in multimodal microscopes where – apart from CARS signals – one is interested in generating and detecting second or third harmonic or two-photon fluorescence [5-11], where pulse durations in the 100-fs range are desirable. A flexible approach consists in introducing a controlled chirp in both beams and increase the duration of the pulses. This can be achieved using pulse stretchers or dispersing elements (e. g. blocks of glass) [12, 13] placed in the paths of ~100-fs pump and Stokes beams. If the chirp is chosen appropriately, the instantaneous frequency difference between the pump and Stokes beam is constant as a function of time. This approach was shown to yield improved spectral selectivity, comparable with that obtainable with transform-limited pulses with the same duration of the chirped pulses, and increased ratio between the resonant and non-resonant signals. It was also exploited in stimulated-Raman scattering [14] and implemented using blocks of glass as stretchers [15] also in single-source CARS microscopy [16], and in broadband CARS [17]. Removing the chirping elements easily restores the sub-ps duration of the pulses if needed. A similarly flexible approach was also proposed in Ref. [18] where only the pump-and-probe beam was strongly chirped. A successful implementation of these techniques requires determining the characteristics of the pulses at the sample level in order to appropriately match their chirps and durations.

Pulses are routinely characterized by means of autocorrelators, where a time-delayed fraction of a pulse interferes on an object having a non-linear optical response with the non-delayed remaining of the same pulse. Second-order processes are recorded as a function of the time delay yielding the beam autocorrelation function and – when the spectral bandwidth and pulse-envelope shape are known – the pulse duration and chirp. Anyway, this characterization cannot be easily performed at the focal plane of the microscope objective and the additional contributions to the total group delay dispersion (GDD) are typically only estimated.

The approach proposed here leads to distinctive advantages even when the contribution of the objective to the total GDD is small with respect to the other optical components in the microscope (or if it can be estimated). In this case the pulse characteristics could be determined with an external autocorrelator by removing the objective and the condenser but this process is cumbersome and can lead to misalignments in the optical path. It should also be noted that optimization strategies common for example in two-photon-excited-fluorescence microscopy, based on maximizing the signal by tuning a pre-chirp device, cannot be applied here. Indeed, the setting that would lead to maximum signal would likely be different to that leading to optimal spectral resolution due to the presence of non-resonant background.

Here a method to determine the features of both the pump-and-probe pulse and the Stokes pulse at the sample without additional autocorrelators is proposed and experimentally verified. The present approach relies on the delay line, ubiquitous in CARS microscopes, to perform a convolution of the pump-and-probe and Stokes pulses as a function of their relative delay $\delta t$ and it is based on the detection of the photons emitted from an appropriate non-linear sample. Noteworthy, when a non-linear sample is available, our method can be completely automated, making pulse-characteristic determination an easy and straightforward procedure. By measuring the intensities of the emitted non-resonant four-wave-mixing (FWM) and sum-frequency-generation (SFG) signals as a function of $\delta t$, the pulse durations at the sample level can be straightforwardly measured. Additionally, if the emitted signals can also be spectrally analyzed, the GDD of each pulse can be directly determined. The only assumptions behind this approach are that: i) the non-linear responses of the chosen medium are only weakly dependent on the wavelength of the excitation photons, at least within the bandwidths of the pulses, which can be easily verified experimentally. ii) Pump and Stokes beam powers are chosen in order to avoid pump depletion. Also this can be checked for experimentally (see for example S2 Fig.). Interestingly, the approach detailed here does not require that one of the pulses is much longer than the other. In fact, by collecting both the FWM and SFG signals also pulses of similar durations and GDDs can be fully characterized. It should be noted here that for the SFG signal to be present, a non-centrosymmetric sample is needed.

## Theory

The electric field of a Fourier-limited Gaussian pulse having duration $\tau$, amplitude $A^{1/2}$, and centered around an angular frequency $\omega$ can be described as:

$$F_{\text{transform limited}}(t) = \sqrt{A} e^{-2t^2 \ln 2/\tau^2} e^{-i\omega t}. \tag{1}$$

After travelling through a dispersive medium the pulse acquires a GDD, quantified here by a parameter *a*. Neglecting third-order dispersion terms and higher, the new pulse can be written as:

$$F(t) = \sqrt{A'} e^{-2t^2 \ln 2 / \tau_{out}^2} e^{-it\omega + 8it^2 a \left( \frac{\ln 2}{\tau \tau_{out}} \right)^2}, \tag{2}$$

where:

$$\tau_{out} = \sqrt{\tau^2 + (8a \ln 2 / \tau)^2} \tag{3}$$

is the pulse duration at the output of the dispersive medium and *A'* the new amplitude.

In the following, we shall refer to the Stokes and pump-and-probe pulses of typical two-color CARS implementations by using "Stokes" and "pump" subscripts in the relevant quantities. The pulses travel through several dispersing elements in the setup, including for example the scan and tube lenses, the microscope objective and any additional glass element introduced to tune their chirp. In the following, *a* will refer to the overall GDD.

The amplitudes of the sum-frequency-generation signal $A_{SFG}$ and of the four-wave mixing one $A_{FWM}$ can be written as:

$$\begin{aligned} A_{SFG}(t, \delta t) &= \sigma_{SFG} F_{pump}(t + \delta t / 2) F_{Stokes}(t - \delta t / 2) \\ A_{FWM}(t, \delta t) &= \sigma_{FWM} F_{pump}(t + \delta t / 2) F^*_{Stokes}(t - \delta t / 2) F_{pump}(t + \delta t / 2), \end{aligned} \tag{4}$$

where $\sigma_{SFG}$ and $\sigma_{FWM}$ describe the process efficiencies, and the "*" superscript indicates the complex conjugate of the field. Here positive time delays $\delta t$ correspond to the pump beam arriving later than the Stokes beam, and we shall take into account only the contribution to the FWM signal generated from the absorption of two pump photons and the stimulated emission of a Stokes one.

Second-harmonic generation (SHG) from the pump pulses and from the Stokes pulses will also be generated. Since these obviously do not depend on $\delta t$, they do not contain cross-correlation terms between the two different pulses and therefore will not be further considered in the following.

The squared moduli of the Fourier transforms of $A_{SFG}$ and $A_{FWM}$ provide their spectra:

$$\tilde{I}_{SFG}(\omega,\delta t) \propto e^{-\frac{16 A_{SFG}^{(0)} \delta t^2 + 8 A_{SFG}^{(1)} \delta t (\omega_{pump}+\omega_{Stokes}-\omega) + A_{SFG}^{(2)} (\omega_{pump}+\omega_{Stokes}-\omega)^2}{4 D_{SFG}}}$$

$$\tilde{I}_{FWM}(\omega,\delta t) \propto e^{-\frac{32 A_{FWM}^{(0)} \delta t^2 - 16 A_{FWM}^{(1)} \delta t (2\omega_{pump}-\omega_{Stokes}-\omega) + A_{FWM}^{(2)} (2\omega_{pump}-\omega_{Stokes}-\omega)^2}{4 D_{FWM}}}.$$

(5)

The explicit expressions of the coefficients in Eq. (5) are reported in S1 Table.

From Eq. (5) it is straightforward to notice that at fixed $\delta t$, the spectra are Gaussian curves with amplitudes ($I_{SFG}$ and $I_{FWM}$) and centers ($\omega_{center,SFG}$ and $\omega_{center,FWM}$) depending on $\delta t$. The amplitudes of the Gaussian curves as a function of $\delta t$ are:

$$I_{SFG}(\delta t) \propto e^{-\frac{4\beta_{pump}\beta_{Stokes}\delta t^2}{\beta_{pump}+\beta_{Stokes}}}$$

$$I_{FWM}(\delta t) \propto e^{-\frac{8\beta_{pump}\beta_{Stokes}\delta t^2}{2\beta_{pump}+\beta_{Stokes}}},$$

(6)

where $\beta_{pump}$ and $\beta_{Stokes}$ are defined in S1 Table. The full-widths at half maximum ($W$) are:

$$W_{SFG}^2 = \ln 2 \left( \frac{1}{\beta_{pump}} + \frac{1}{\beta_{Stokes}} \right)$$

$$W_{FWM}^2 = \ln 2 \left( \frac{1}{2\beta_{pump}} + \frac{1}{\beta_{Stokes}} \right).$$

(7)

The FWHMs can be straightforwardly determined experimentally by monitoring $I_{SFG}$ and $I_{FWM}$ while moving the delay line. Inverting Eq. (7) allows calculating the values of $\tau_{out,pump}$ and $\tau_{out,Stokes}$:

$$\tau_{out,pump}^2 = 2(W_{SFG}^2 - W_{FWM}^2)$$

$$\tau_{out,Stokes}^2 = 2W_{FWM}^2 - W_{SFG}^2.$$

(8)

If the pulse durations are known, then the inversion of Eq. (3) yields the GDDs at sample level. If the durations are not known but the center frequency of the SFG and FWM signals can be determined by acquiring their spectra, it is still possible to retrieve the values of the GDDs. To this end, it should be noted from Eq. (5) that the center frequencies of the spectra shift linearly as a function of $\delta t$ with rates ($S$):

$$S_{SFG} \equiv \frac{d\omega_{center,SFG}}{d\delta t} = 4\frac{\beta_{pump}\alpha_{Stokes} - \alpha_{pump}\beta_{Stokes}}{\beta_{pump} + \beta_{Stokes}}$$
$$S_{FWM} \equiv \frac{d\omega_{center,FWM}}{d\delta t} = -8\frac{\beta_{pump}\alpha_{Stokes} + \alpha_{pump}\beta_{Stokes}}{2\beta_{pump} + \beta_{Stokes}}. \tag{9}$$

Inversion of Eq. (9), after measuring the rates and calculating the values of $\tau_{out,pump}$ and $\tau_{out,Stokes}$, yields the values of $\alpha_{pump}$ and $\alpha_{Stokes}$:

$$\alpha_{pump} = -\frac{W_{FWM}^2 S_{FWM} + W_{SFG}^2 S_{SFG}}{8\tau_{out,pump}^2}$$
$$\alpha_{Stokes} = \frac{W_{SFG}^2 S_{SFG} - W_{FWM}^2 S_{FWM}}{8\tau_{out,Stokes}^2}, \tag{10}$$

and, straightforwardly, of $\tau_{pump}$, $\tau_{Stokes}$, $a_{pump}$, and $a_{Stokes}$:

$$\tau_{pump}^2 = \frac{\beta_{pump}\ln 2}{\alpha_{pump}^2 + \beta_{pump}^2}$$
$$\tau_{Stokes}^2 = \frac{\beta_{Stokes}\ln 2}{\alpha_{Stokes}^2 + \beta_{Stokes}^2}$$
$$a_{pump} = \frac{\alpha_{pump}}{4\left(\alpha_{pump}^2 + \beta_{pump}^2\right)}$$
$$a_{Stokes} = \frac{\alpha_{Stokes}}{4\left(\alpha_{Stokes}^2 + \beta_{Stokes}^2\right)}, \tag{11}$$

yielding the complete characterization of the pulses at the sample level.

## Materials and Methods

In the following, we shall show the validity of our approach by reporting the pulse characteristics of a CARS setup, schematically shown in Fig. 1a, based on a fs laser for the generation of the pump-and-probe pulses with a wavelength of 810 nm and an OPO for the Stokes pulses (1060-nm wavelength). The beams propagate through two blocks of SF6 glass and two beam expanders. In the following we shall indicate with $L_{pump}$ ($L_{Stokes}$) the length of the glass blocks placed in the pump (Stokes) path. Before the dichroic mirror D, the Stokes pulses are delayed by means of a motorized delay line. Photons reach and overfill the pupil of the microscope objective after being reflected by a pair of galvanometric mirrors and after propagating through a scan lens and a tube lens. Pump (Stokes) beam power measured after the objective is 20 mW (10 mW). Signals generated from the sample are collected by the microscope condenser, spectrally filtered to remove pump and Stokes photons, and analyzed by a spectrometer.

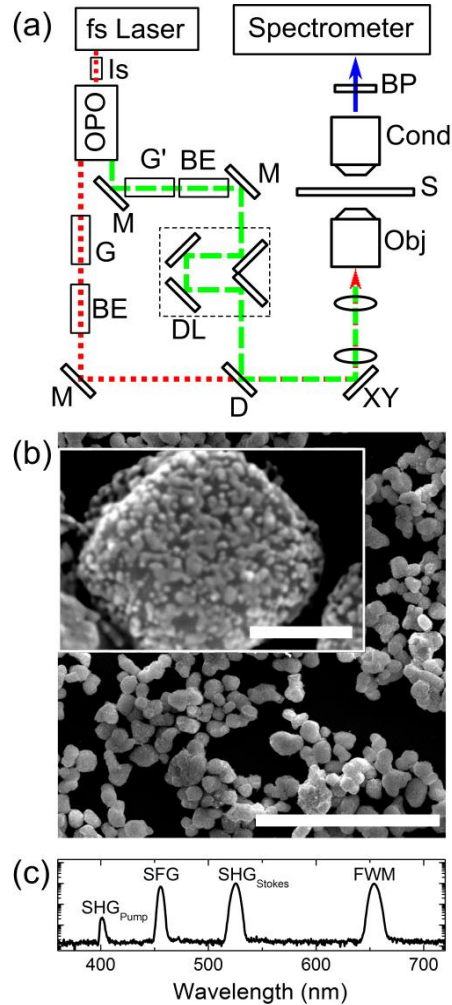

FIG. 1. a) Schematic view of the CARS microscopy setup used in this work. Legend: fs laser (Chameleon Vision II). Is: Faraday optical isolator. OPO: Radiantis ORIA optical parametric oscillator. G, G': dispersing SF6-glass blocks. BE: beam expanders. M: mirrors. DL: delay line. D: dichroic mirror. XY: galvanometric mirrors. Obj: Zeiss 32x N.A.=0.85 C-Achroplan. S: sample. Cond: microscope condenser. BP: razor edge and band pass filters. Spectrometer: Jobin Yvon HR550 monochromator coupled to a Synapse CCD camera. The pump-and-probe (Stokes) beam path is shown as dotted (dashed) lines. b) Scanning-electron-microscope image of core/shell of the $BaTiO_3$/Au nanoparticles used in this work (scale bar: 5 μm). Inset: magnified view of an individual nanoparticle showing decoration of the core with Au clusters (scale bar: 200 nm). c) Semi-log plot of the spectrum collected from a nanoparticle without G and G' and with the pump-and-probe and Stokes beam overlapping in time.

As a sample material for the experiments, we chose core/shell $BaTiO_3$/Au nanoparticles, (NPs), with an average diameter of 300 nm owing to their strong non-linear optical properties. A scanning-electron-microscope image of some NPs deposited on a microscope glass slide is shown in Fig. 1b. The full synthesis and characterization details of the NPs are reported elsewhere [9,19].

## Results and Discussion

In order to test our approach, we introduced known amounts of GDD in the pump and Stokes paths, determined the characteristics of the pulses, and checked them for consistency. The first test was carried out in the absence of SF6 glasses ($L_{pump} = L_{Stokes} = 0$) to evaluate the contribution of the set-up optics and sources alone to the total GDDs. The measured spectra as a function of the delay are reported in Fig. 2a, where the top left (bottom left) panel shows the SFG (FWM) signal. Each spectrum was fitted with a Gaussian curve, yielding its center wavelength (shown in Fig. 2a, left panels, as dashed curves, as a function of the delay) and its amplitude (shown in Fig. 2b as open circles). The latter was fitted as a function of the delay again with Gaussian curves yielding $W_{SFG}$ and $W_{FWM}$. Linear fits of the center wavelengths as a function of the delay in turns gives $S_{SFG}$ and $S_{FWM}$. Equations (8), (10), (11) finally gives the pulse parameters, which were used to calculate the theoretical spectra (Fig. 2a, right panels) for comparison with the experimental data.

The parameters calculated for $L_{pump} = L_{Stokes} = 0$ and for the other configurations tested, averaged over 20 repetitions of each experiment, are reported in S2 Table and in S1 Fig. It is interesting to note that, without G and G' (i. e. $L_{pump} = L_{Stokes} = 0$) even if both pulses cross similar amounts of glass in the setup, they reach the sample with a difference in GDD ~10000 $fs^2$. This is quite likely due to a setting of the OPO cavity introducing a negative chirp in the Stokes beam that compensates the positive one introduced by the additional optics.

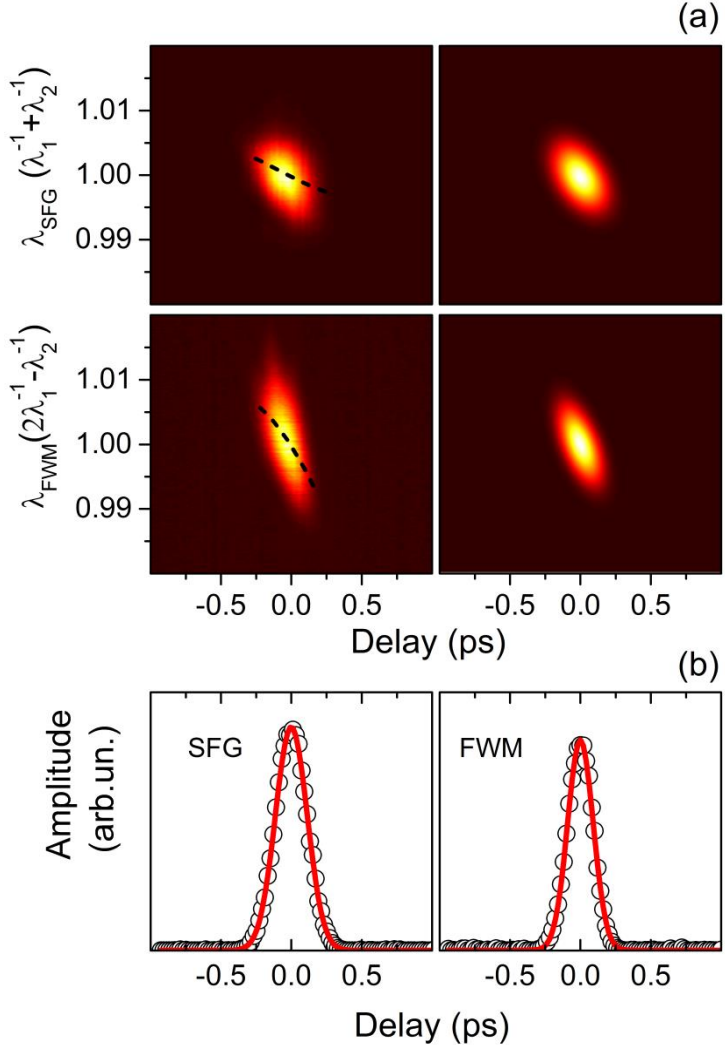

FIG. 2. a) Measured (left panels) and simulated (right panels) sum-frequency generation (top row) and four-wave mixing (bottom row) signals with $L_{pump} = L_{Stokes} = 0$. Data and calculation results are shown normalized from 0 (dark) to 1 (white). The dashed lines represent the centers of the spectra determined by Gaussian fits of the data. b) Amplitude of the SFG (left) and FWM (right) spectra shown in Panel (a) as a function of the delay. Experimental data are shown as circles. Gaussian fits to the data are displayed as solid lines.

We repeated the procedure after introducing $L_{pump}$ = 10 cm and $L_{Stokes}$ = 15 cm blocks of SF6 glass. The lengths were calculated in order to introduce approximately the same GDD for both the pump and Stokes beams, which is what one would be tempted to choose without knowing the real characteristics of the pulses. Calculating the refractive index of the SF6 glass by means of the Sellmeier equation (with parameters $B_1$ = 1.72448482, $B_2$ = 0.390104889, $B_3$ = 1.04572858, $C_1$ = 0.0134871947 µm$^2$, $C_2$ = 0.0569318095 µm$^2$, $C_3$ = 118.557185 µm$^2$), the additional dispersions can be determined to be 19643 fs$^2$ and 19746 fs$^2$ for the pump and Stokes pulses respectively. The characterization of SFG and FWM, shown in Fig. 3, yielded pulse durations – reported in S2 Table and in S1 Fig. – consistent with the previous configurations and also GDD values in agreement with those determined for $L_{pump} = 0$ and $L_{Stokes} = 0$, increased by the calculated GDDs introduced by the glass. Similarly consistent parameters were obtained for $L_{pump} = 0$ cm and $L_{Stokes} = 15$ cm.

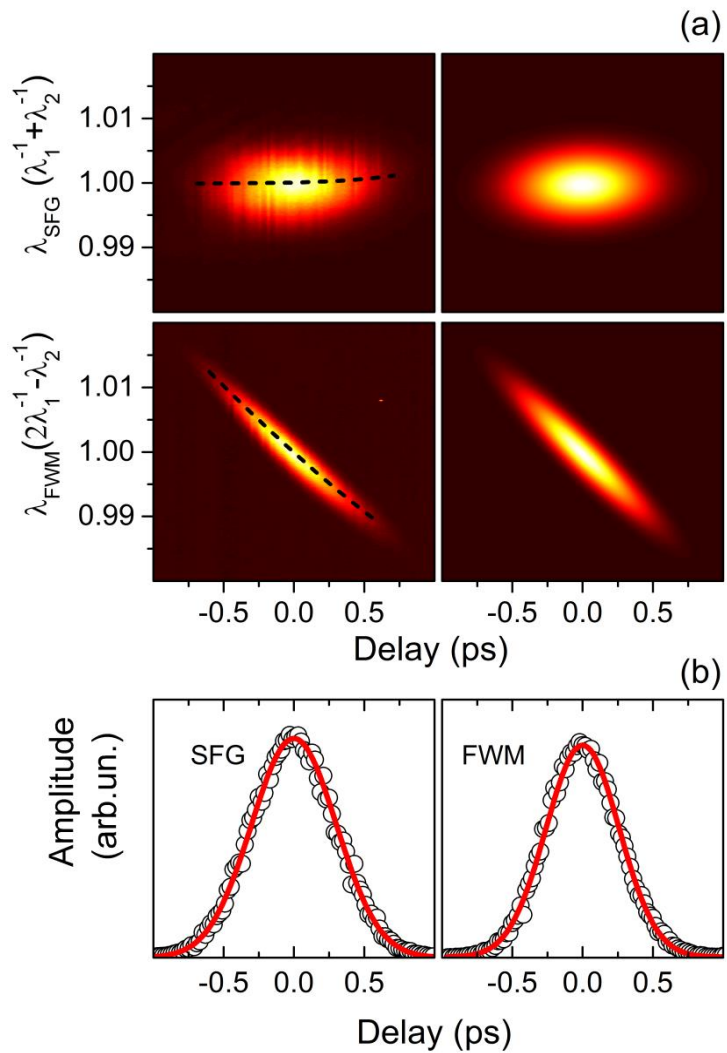

FIG. 3. a) Measured (left panels) and simulated (right panels) sum-frequency generation (top row) and four-wave mixing (bottom row) signals with $L_{pump} = 10$ cm, $L_{Stokes} = 15$ cm. Data and calculation results are shown normalized from 0 (dark) to 1 (white). The dashed lines represent the centers of the spectra determined by Gaussian fits of the data. b) Amplitude of the SFG (left) and FWM (right) spectra shown in Panel (a) as a function of the delay. Experimental data are shown as circles. Gaussian fits to the data are displayed as solid lines.

The last combination of glass blocks used, namely $L_{pump}$ = 10 cm and $L_{Stokes}$ = 25 cm, was determined following the characterization in the other configurations in order to yield similar GDDs (~30000 fs$^2$) for both pulses at the sample level. The increase in spectral resolution with the latter choice of glass lengths was demonstrated by measuring the spectrum of liquid methanol in the C-H stretch region (shown in Fig. 4) and comparing it with the spectrum calculated using the pulse characteristics determined with our algorithm. The height of the Raman peaks and the amount of non-resonant background were adjusted to yield the best fit to the data, while their positions were fixed at 2944 cm$^{-1}$ and 2836 cm$^{-1}$, with widths of 34 and 21 cm$^{-1}$ respectively, according to Ref. [20]. Figure 4 also shows the calculated spectra corresponding to $L_{pump}$ = 10 cm and $L_{Stokes}$ = 15 cm and to $L_{pump}$ = 0 and $L_{Stokes}$ = 0. The spectrum calculated with the pulse parameters determined with the analysis presented here provides a very good match to the experimental data, further confirming the validity of our approach.

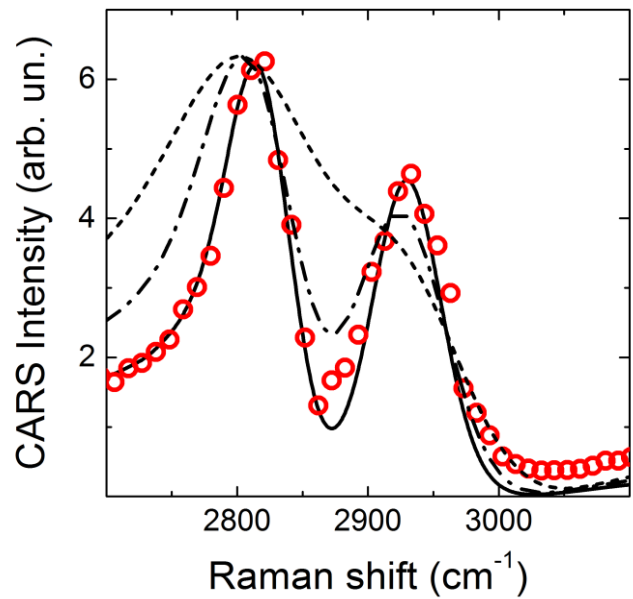

FIG. 4. CARS spectra of liquid methanol. The experimental spectrum, measured with $L_{pump}$ = 10 cm and $L_{Stokes}$ = 25 cm is shown with circles. The corresponding calculated spectrum is shown as a black solid line. Spectra calculated for $L_{pump}$ = 0 and $L_{Stokes}$ = 0 (dashed line) and $L_{pump}$ = 10 cm and $L_{Stokes}$ = 15 cm (dash-dotted line) are also reported for comparison.

# Conclusions

We have shown that the pulse characteristics of a typical CARS setup can be easily retrieved, without using additional instrumentation, by analyzing the FWM and SFG signals from an appropriate non-linear sample. This information can be exploited to precisely determine and tune the chirp and pulse duration at sample level and therefore to optimize the performance of CARS microscopes in terms of spectral resolution and imaging contrast, and signal level for each specific application.

# Supporting Information

**S1 Table. Explicit expressions of the coefficients displayed in Eq. (5).**

| | |
|---|---|
| $A_{SFG}^{(0)}$ | $\beta_{pump}\alpha_{Stokes}^2 + \alpha_{pump}^2\beta_{Stokes} + \beta_{pump}\beta_{Stokes}^2 + \beta_{pump}^2\beta_{Stokes}$ |
| $A_{SFG}^{(1)}$ | $\beta_{pump}\alpha_{Stokes} - \alpha_{pump}\beta_{Stokes}$ |
| $A_{SFG}^{(2)}$ | $\beta_{pump} + \beta_{Stokes}$ |
| $D_{SFG}$ | $\left(\alpha_{pump} + \alpha_{Stokes}\right)^2 + \left(\beta_{pump} + \beta_{Stokes}\right)^2$ |
| $A_{FWM}^{(0)}$ | $\beta_{pump}\alpha_{Stokes}^2 + 2\alpha_{pump}^2\beta_{Stokes} + \beta_{pump}\beta_{Stokes}^2 + 2\beta_{pump}^2\beta_{Stokes}$ |
| $A_{FWM}^{(1)}$ | $\beta_{pump}\alpha_{Stokes} + \alpha_{pump}\beta_{Stokes}$ |
| $A_{FWM}^{(2)}$ | $2\beta_{pump} + \beta_{Stokes}$ |
| $D_{FWM}$ | $\left(2\alpha_{pump} - \alpha_{Stokes}\right)^2 + \left(2\beta_{pump} + \beta_{Stokes}\right)^2$ |
| $\alpha_i$ | $4a_i\left(\tau_i\,\tau_{\text{out},i}\right)^{-2}\ln^2 2 \qquad i \in (pump, Stokes)$ |
| $\beta_i$ | $\tau_{\text{out},i}^{-2}\ln 2$ |

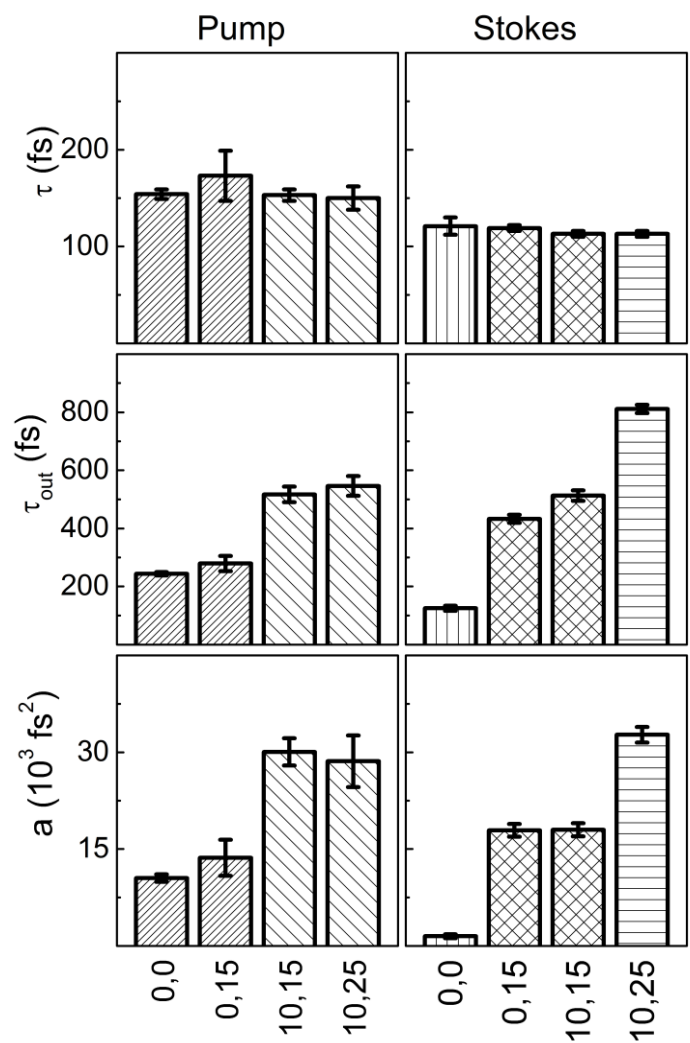

S1 Fig. Summary of the pump-and-probe and Stokes pulse characteristics with different lengths of SF6 glass introduced in the beam paths. The labels on the horizontal axis represent the lengths in cm of the glass introduced in the pump-and-probe path and in the Stokes one. Bars are pattern-filled depending on the amount of glass introduced in the pump-and-probe path (left column) or in the Stokes path (right column) to allow an easy comparison among the data. Error bars show the standard deviation calculated over 20 repetition of the acquisition.

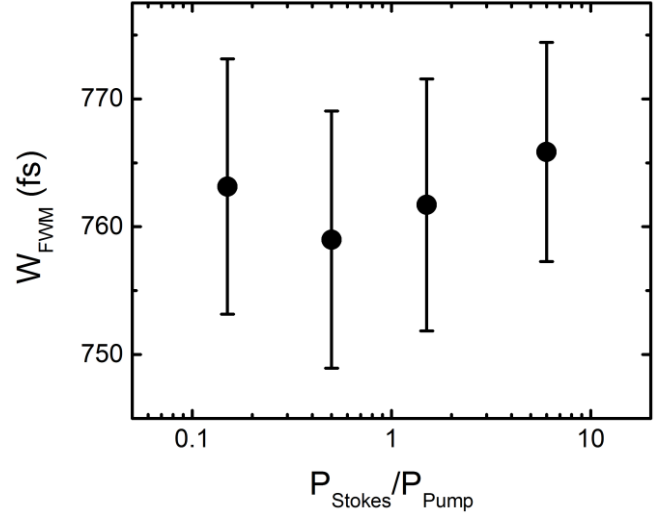

S2 Fig. Measured values of $W_{FWM}$ as a function of the ratio between the Stokes and pump laser powers with $L_{pump}$ = 10 cm and $L_{Stokes}$ = 25 cm. Within the measurement uncertainties, no variations of $W_{FWM}$ were observed. In the case of probe-induced pump depletion, at large ratios a reduction of $W_{FWM}$ is expected.

S2 Table. Pulse parameters determined as a function of $L_{pump}$ and $L_{Stokes}$. Errors represent the standard deviation calculated over 20 repetition of the acquisition.

| | | | | |
|---|---|---|---|---|
| $L_{pump}$ (cm) | 0 | 0 | 10 | 10 |
| $L_{Stokes}$ (cm) | 0 | 15 | 15 | 25 |
| $\tau_{out,pump}$ (fs) | 244 ± 6 | 280 ± 25 | 520 ± 17 | 545 ± 35 |
| $\tau_{out,Stokes}$ (fs) | 125 ± 10 | 435 ± 15 | 515 ± 20 | 810 ± 14 |
| $\tau_{pump}$ (fs) | 155 ± 5 | 170 ± 25 | 153 ± 6 | 150 ± 12 |
| $\tau_{Stokes}$ (fs) | 120 ± 10 | 120 ± 3 | 113 ± 3 | 113 ± 3 |
| $a_{pump}$ (fs$^2$) | 10500 ± 600 | 13700 ± 2280 | 30000 ± 2100 | 28600 ± 4000 |
| $a_{Stokes}$ (fs$^2$) | 1500 ± 300 | 17900 ± 1000 | 18000 ± 1000 | 32700 ± 1200 |